# Architected Materials for Mechanical Compression: Design via Simulation, Deep Learning, and Experimentation

Andrew J. Lew[1,2], Kai Jin[1], Markus J. Buehler[1,3*]

[1] Laboratory for Atomistic and Molecular Mechanics (LAMM), Massachusetts Institute of Technology, 77 Massachusetts Ave., Cambridge, MA 02139, USA

[2] Department of Chemistry, Massachusetts Institute of Technology, 77 Massachusetts Ave., Cambridge, MA 02139, USA

[3] Center for Computational Science and Engineering, Schwarzman College of Computing, Massachusetts Institute of Technology, 77 Massachusetts Ave., Cambridge, MA 02139, USA

[*] email: mbuehler@mit.edu

**Abstract**

Architected materials can achieve enhanced properties compared to their plain counterparts. Specific architecting serves as a powerful design lever to achieve targeted behavior without changing the base material. Thus, the connection between architected structure and resultant properties remains an open field of great interest to many fields, from aerospace to civil to automotive applications. Here, we focus on properties related to mechanical compression, and design hierarchical honeycomb structures to meet specific values of stiffness and compressive stress. To do so, we employ a combination of techniques in a singular workflow, starting with molecular dynamics simulation of the forward design problem, augmenting with data-driven artificial intelligence models to address the inverse design problem, and verifying the behavior of *de novo* structures with experimentation of additively manufactured samples. We thereby demonstrate an approach for architected design that is generalizable to multiple material properties and agnostic to the identity of the base material.

**Introduction**

Hierarchical materials with specific architecture at different length scales are observed everywhere in nature[1], like in bone[2] and wood.[3] Adding architecture to structures can enhance mechanical properties and provides an extra design lever on top of atomic-level microstructure and macroscopic-level part dimensions.[1,4–6] Investigations into hierarchically architected materials have thus been of great interest, with efforts to control fatigue tolerance,[7] energy absorption,[8] and stiffness and strength,[9] among many others. Design approaches for architected materials have included inspiration from existent crystalline material microsctructure,[10] finite element topology optimization,[11] and experimentation with rapidly fabricated additively manufactured samples.[12] Honeycomb structures are of particular interest due to their ultra-low weight and outstanding mechanical properties, with a variety of applications across automotive, railway, and aerospace industries.[13]

Recent advances in artificial intelligence have afforded new capabilities for architectural design. For example, there have been successes in achieving bioinspired hierarchical composites,[14] in using semi-supervised approaches with graph neural networks,[15] and in implementing natural language inputs for generative design[16–20] of architected materials. Concurrently, machine learning models have been used in other material platforms for the prediction of a multitude of mechanical properties including fracture,[21–23] compliance,[24] and buckling.[25,26]

Here we demonstrate a full workflow to tackle compression design of architected honeycomb materials, utilizing simulation to determine initial insights into the space of hierarchical honeycomb lattices, machine learning and genetic algorithms to generate candidates for desired behavior, and additive manufacturing to rapidly test top structural candidates. This approach is generalizable to multiple material property targets, with 4 levels each of stiffness and stress illustrated here.

**Results**

We simulate the compressive behavior of honeycomb structures as in Figure 1a via a coarse-grained molecular dynamics approach,[27] in which we represent the honeycomb lattice as a series of interconnected beads subject to harmonic bond interactions, harmonic angle interactions, and a Lennard-Jones potential. While alternatively, 2D finite element beams could be used to model these honeycomb structures, molecular dynamics was used because it can better handle a variety of more complex, non-beam-based structures and could be easily adapted to deal with various nanoscale materials in the future. Hence, this strategy provides a flexible foundation to the approach that increases its applicability to more general cases. We then define an initial set of hierarchical honeycomb structures. These structures are comprised of a regular honeycomb lattice, onto which an additional longer-range hexagonal lattice is superimposed. Specifically, this superlattice is formed by adjusting the thicknesses of cell walls in a higher order hexagonal pattern. Here we vary both the repeat length of the superlattice and the cell wall thicknesses of each hexagonal ring within the superlattice to obtain a variety of structures with diverse compressive behaviors, as shown in Figure 1b. These super-honeycomb structures are encoded as vectors, in which each entry of the vector dictates the relative thickness of the corresponding hexagonal arrangement of cell walls within the superlattice. Stress curves are natively representable as vectors, and are sliced into progressive time series in order to provide sections of compressive behavior at different timesteps.

Formatting structure and stress data in these ways allows us a natural way to map a cause and effect relationship between structure and property. Specifically, we append snapshot stress states to each structure vector, and pair off each such appended vector with its subsequent window of stress evolution. We then train a machine learning model to learn the relationship between a structure at a particular stress state as input and its subsequent stress behavior for the upcoming time window as output, as shown in Figure 1c. In Figure 1d, the final trained model can be deployed to make full predictions of the stress strain curve, starting from an initial unstressed structure. A zoomed in image of a hierarchical honeycomb structure is shown, to illustrate how encoded structural parameters correspond to the geometry. Details of MD simulation and data representation are provided in the Methodology section.

Over 1,000 MD simulations with randomized superlattice sizes and cell wall thicknesses, sliced into over 26,000 input-output training pairs, provide a training dataset with diverse stress behaviors, as shown in Figure 2a. Training a convolutional LSTM model, suitable for learning time dependent data such as stress progression,[28] yields accurate predictions of compression behavior. Figure 2b compares real versus predicted stress values and illustrates a linear fit with an $r^2$ = 0.95. Figure 2c shows good agreement between training and validation losses, indicating the model is not simply overfitting on the training data, with a validation loss of 0.00058. Using the trained model to predict stress curves of the four samples presented in Figure 1b shows a good ability to accurately predict a wide variety of compressive stress behaviors. Further details of ML model and training are provided in the Methodology section.

We subsequently leverage this model to tackle the inverse design problem, in which we start out with a desired material property and predict a structure that exhibits it. Specifically, we employ an iterative genetic algorithm approach as in our previous work[21], which alternates between stages of structure generation and evaluation, with specific steps outlined in Figure 3 and using structure encodings as the genes subject to evolution.

Briefly, an initial population of structures is randomly generated. Then, two structures are randomly chosen to serve as parents for new children structures, which have some combinations of features similar to the parent structures. Additional sources of structure novelty are introduced through the mutation, specialization, and migration of new structures, in which added structures are either randomly perturbed from existing structures, accentuated in features from existing structures, or newly instantiated completely randomly, respectively.

At the end of this multistep generative process, we have an enlarged population of structures to evaluate. To do this, we first use the ML model to rapidly characterize each structure's compressive stress response. Next, we rank each structure's behavior by a chosen fitness function. For example, we could calculate the initial slopes of each stress strain curve to extract values for material stiffness, and compare these calculated values to a desired target value. Then, proximity to the target value is used as the fitness function. Finally, we screen the ranked structures to reduce the population back to its initial size before moving on to the generation stage in the next iteration of the cycle.

Once structure fitness reaches a threshold value, we break out of the loop and select the best structures from the evaluated population. These structures are identified as likely candidates for possessing the mechanical properties we desire. To increase confidence in their properties, we subject these candidate structures to MD simulation. Figure 4 shows a selection of 8 candidate structures designed for a range of stiffness and ultimate stress values, with good agreement between ML predictions and MD simulations.

After this corroboration with MD, we experimentally verify that candidate structures indeed possess desired properties. To do this, we use additive manufacturing to rapidly fabricate the candidate super-honeycombs out of thermoplastic polyurethane (TPU) and subsequently subject them to compression tests. Figure 5a illustrates four candidate structures designed to have a specific progression of stiffness values, and their additively manufactured physical instantiations subjected to compression testing. The resultant force displacement curves in Figure 5b initially overlap, likely representing the shared response of the base printing material. However, the curves diverge after 10 mm with the onset of different buckling behavior across the different architectures. Starting from this point, comparison with MD shows experimental compressive behavior qualitatively aligns with simulated behavior for the following 10 mm of displacement. Calculated stiffnesses from this divergence point, as the collection of slopes from 8 mm to 13 mm in 3 mm increments, provide a relative progression of values that align well with the desired targets, with error bars corresponding to the standard deviation of these slopes. A comparison plot showing good agreement between target stiffness, ML predictions, MD corroboration, and experimental verification is provided in Figure 5c.

We perform similar experimental verification for structures designed to have specific maximum stress values. Like before, we fabricate structures via additive manufacturing and subject them to compression testing in Figure 6a. Furthermore, the 10 to 20 mm region of force displacement curves again qualitatively correspond to simulated behavior with error bars corresponding to the standard deviation of values within a 5 mm range centered about 20 mm, as shown in Figure 6b. And quantitatively, experimental maximum force values within this region correspond well with ultimate stress values of desired targets, ML predictions, and MD corroboration in Figure 6c.

**Conclusions**

The trained ML model provides an effective tool for the forward design problem, in which a given super-honeycomb structure can have its compressive behavior directly and rapidly predicted without having to set up, run, and analyze a physics-based simulation. A genetic algorithm search validated by simulation and experimentation enables effective interrogation of the inverse design problem.

It should be noted that stress curve predictions were ultimately based on a simplified harmonic Lennard-Jones model without any specific information on what base materials would be used to fabricate structures. It is noted that the coarse grain model is not tuned to replicate the specific performance of only one material, such as the TPU used in this study, as our approach may be of interest to many different material platforms. Even though deformation in experiment is conducted at slow rates, resultant discrepancies, such as experimental force displacement curves corresponding to predictions after 10 mm, are not unexpected. No doubt, if one were particularly interested in honing in on some selected base material, a trained model specifically incorporating data on the unique properties of that particular material would outperform the results here. This is also why the comparisons between computation and experiment are presented in relative rather than absolute terms - one could easily instantiate structures that are stiffer in an absolute sense if one simply used a stiffer polylactic acid (PLA) filament instead of a flexible TPU filament, of course.

However, and most importantly, even in the absence of such specific material property information, the predictions made here align to the crucial time steps around when buckling initiates in terms of both qualitative behavior and overall trends. The rankings for highest to lowest stiffness or ultimate stress remain consistent, as do the quantitative relative distances between values, per Figures 5c and 6c. Indeed, the aim here was not to create a specific model tailored to precisely predict the properties of TPU structures, but rather to tackle the challenge of architectural design in a more general material agnostic manner. If one is interested in designing for more granular scale-dependent localizations, future work may be interested in training the deep convolutional LSTM model on explicit representations of them. Other than mechanical properties, thermodynamics and kinetic theories can also be considered as CG models.[27] Hence, other material properties such as thermal conductivity may be treated by an ML inverse design approach, given a proper training dataset. This provides one of out many avenues for future directions that could build on this work.

Aside from training the predictive model on parameters of real material properties for better consistency with specific experimental cases, further space for advancement lies in the direction of structure generation. For example, future work may be done in modeling more direct, bidirectional translation between structure and property. In a literal sense, the growing field of machine translation[29] of languages may find great utility here, if we treat structural encodings as "words in one language" that should translate to equivalent stress vector "words in a second language". A model fluent in these two languages would be a valuable tool to both forward and inverse design tasks. And a model that gains some understanding of semantic rules and forbidden statements in either "language" may be helpful in putting bounds on the possible structures or properties that can be attained.[17,18]

Another recent development in state-of-the-art image generation concerns diffusion models like DALL·E, which have found great success in generating images based on complex prompts or conditioning parameters.[18,30–32] It may be fruitful to train diffusion models that learn how to architect structures based on prompts detailing mechanical property requirements.

Regardless of such advancements, which may increase the performance of our approach if they were to pan out, here we demonstrate our approach is a successful end-to-end process for compression design from ideated property requirements to actualized material structures. Furthermore, it is not difficult to imagine how such a process can be implemented at scale, with other architecture platforms, and for other properties of interest. We look forward to how the synergy of simulation, artificial intelligence, and experiment can empower materials design in the years to come.

## Methods

**Stress simulation by molecular dynamics.** Structures are encoded as 10-D vectors, where each entry corresponds to the cell wall thicknesses of successive hexagonal rings in the superlattice, as shown in Figure 1d. As we simply join adjacent hexagonal cells along their cell walls, strut thicknesses are determined as the sum of the two adjacent cell thicknesses. For superlattices less than 10 hexagonal units in circumradius, the remaining entries in the 10-D vector are set to zero.

To generate an initial set of data, we randomly instantiate a set of 1,500 structures. For each structure, we select a random number from 2 to 10 from a uniform distribution to serve as the superlattice length. Then, each entry in the superlattice is assigned a random relative cell thickness from 2 to 15, also from a uniform distribution. These structural parameters are used to generate coarse-grained representations of super-honeycomb structures, wherein each structure is comprised of 14,309 coarse-grain beads arranged in a hexagonal lattice, strut thicknesses are normalized to ensure consistent mass across all samples and isolate the effect of material redistribution, and the bead masses and interaction coefficients are adjusted to emulate the behavior of thicker or thinner walls.

First, the average strut thickness A of the honeycomb is obtained by the following equation 1:

$$A = \sum_{i=1}^{N} T_i \cdot L_i / \sum_{i=1}^{N} L_i \qquad (1)$$

for $N$ struts with thicknesses $T_i$ and lengths $L_i$. Then, the strut thicknesses $T_i$ are divided by the average $A$ to yield a new set of normalized thicknesses $t$ that yield a structure with an average thickness of 1. We then use the normalized thicknesses t to determine bead masses m as in the following equation 2:

$$m = \rho r_0 t \qquad (2)$$

with density ρ, and equilibrium bead distance $r_0$ set to 5 Å. Then the harmonic bond interactions $U_b$ are determined by the following equation (3):

$$U_b = \frac{Et}{2r_0}(r - r_0)^2 \qquad (3)$$

for elastic modulus E set to 100 MPa. Similarly, the harmonic angle interactions $U_a$ are determined by the following equation (4):

$$U_a = \frac{Et^3}{24r_0}(\theta - \theta_0)^2 \qquad (4)$$

with equilibrium bead angle $\theta_0$ set to 180° within the struts and to 120° at the vertices. We add a Lennard-Jones pair style potential with a cutoff of 2.5 nm to prevent particle penetration.

These structures are then subject to compression testing with MD simulations using LAMMPS,[33] with initial equilibration using the microcanonical ensemble (NVE) with additional temperature rescaling to maintain 10 K during a brief finite temperature equilibration phase before energy minimization. The structures all share the same width, height, mass, and thus density. The difference between structures is in the distribution of this mass within the designated sample area. The equilibration and compression conditions are kept consistent between samples to isolate the effect of varying structure by material redistribution. Compression is implemented by iteratively deforming the simulation box and performing energy minimization via the Polak-Ribiere version of the conjugate gradient algorithm at each increment. For each structure, we obtain a stress strain curve, which we then slice into progressive sequences that highlight evolution over time.

The stress strain relation is naturally embedded in the outcome of the coarse grain MD simulations, obtained via LAMMPS compute pressure, as the sum of contributions from bond, angle, and pair interactions. As our focus lay in designing stress strain behavior of each structure as a whole, we did not systematically observe local scale-dependencies. Of the 1,500 generated structures, 1,445 full stress strain curves are obtained, as some structures fail during compression and yield incomplete stress curves.

The final format for simulation data, amenable for subsequent training of a machine learning model, is paired up in relationships of input-output. Input data are comprised of 13-D vectors, 10 of which correspond to the previously described structure encoding and 3 of which correspond to the current strain step, current stress state, and next stress state of the structure, respectively. Output data are comprised of 4-D vectors representing the subsequent 4 stress steps for a given input. In total, we obtain 34,680 input-output pairs after slicing each original stress curve into stress progressions from 24 different timesteps.

The coarse grain beads have an equilibrium spacing of 5 Å, compared to a simulation sample width of 173 nm. When translated to the printed samples, which are scaled up to a sample width of 100 mm, this corresponds to a coarse grain size of 289 μm.

**Stress prediction by machine learning.** We use a deep convolutional LSTM model[34] suitable for learning temporal relationships between data. Training is done using an Adam optimizer[35] with a learning rate of 0.0001, decay of 0.001, and batch size of 32. 75% of the MD data is used for training, while 25% is reserved for validation. The specific layer architecture used is described in Table 1.

**Table 1.** Summary of the ML model architecture.

| Layer | Output Shape | Parameters |
| --- | --- | --- |
| Dense | (None, 13) | 182 |
| Dense | (None, 100) | 1400 |
| Reshape | (None, 25, 4) | 0 |
| Conv1D | (None, 25, 4) | 52 |
| BatchNormalization | (None, 25, 4) | 16 |
| Conv1D | (None, 25, 16) | 208 |
| BatchNormalization | (None, 25, 16) | 64 |
| Conv1D | (None, 25, 64) | 3136 |
| BatchNormalization | (None, 25, 64) | 256 |
| Conv1D | (None, 25, 128) | 24704 |
| BatchNormalization | (None, 25, 128) | 512 |
| Conv1D | (None, 25, 64) | 24640 |
| BatchNormalization | (None, 25, 64) | 256 |
| Conv1D | (None, 25, 16) | 3088 |
| BatchNormalization | (None, 25, 16) | 64 |
| Conv1D | (None, 25, 4) | 196 |
| BatchNormalization | (None, 25, 4) | 16 |
| LSTM | (None, 512) | 1058816 |
| Dense | (None, 4) | 2052 |

We use an iterative process similar to our previous work with LSTM models[22,25] for obtaining total stress predictions. Here, we use previous predictions to update the "current" stress and strain values used for subsequent predictions. As 4 steps of prediction are output for each input, and we advance forward inputs 1 timestep at a time, we save a running list of up to 4 potential predicted stress values for each timestep. Once we reach a given timestep and need to use it as input for the next prediction, we average all previous output predictions for that timestep and finalize this as the most likely current value. We continue this process until all desired timesteps are filled.

**Rapid fabrication by additive manufacturing.** 2D images of candidate honeycomb structures were extruded into 100 mm x 67.17 mm x 22.4 mm blocks and sliced to .3mf files with PrusaSlicer 2.5.0. Samples were printed on an Original Prusa i3 MK3 using a 1.75mm green NinjaFlex TPU filament.

**Experimental verification by compression testing.** Compression testing was performed using a 5 kN Instron Universal Testing Instrument. Compression tests were conducted at a rate of 10 mm/min to a final displacement of 30 mm. Videos of samples under compression were taken with a Samsung Galaxy Note20 Ultra 5G 108 MP wide-angle camera. We do not expect a direct quantitative correspondence between simulated and experimental stress values due to the omission of material specific parameters in the coarse grain MD simulations. To avoid potential confusion of conflating material-agnostic values from coarse grain MD simulation with the specific values from TPU experimentation, we do not represent the two sets of curves as direct comparisons of stress and strain. Rather, the experimental compression tests are provided in their native load displacement format, and comparisons are made by relative trends.

## Authors' Contributions

A.J.L. defined the distribution of structures to serve as training data, ran the MD simulations, made and trained the ML model, implemented the genetic algorithm, 3D printed the samples, conducted experimental compression testing, and wrote the manuscript. K.J. wrote the codes to automate generation of molecular dynamics scripts from structural parameters and to produce .STL files from structural images. M.J.B. oversaw all work, designed the ML model with A.J.L, analyzed data with A.J.L., and edited the manuscript.

## Acknowledgements

This material is based upon work supported by the NSF GRFP under Grant No. 1122374. We acknowledge support by NIH (5R01AR077793-03), the Office of Naval Research (N000141612333 and N000141912375), AFOSR-MURI (FA9550-15-1-0514) and the Army Research Office (W911NF1920098). Related support from the MIT-IBM Watson AI Lab, MIT Quest, and Google Cloud Computing, is acknowledged.

## Data Availability

The data that support the findings of this study are available from the corresponding author upon reasonable request.

**Competing Interests**

The Authors declare no Competing Financial or Non-Financial Interests.

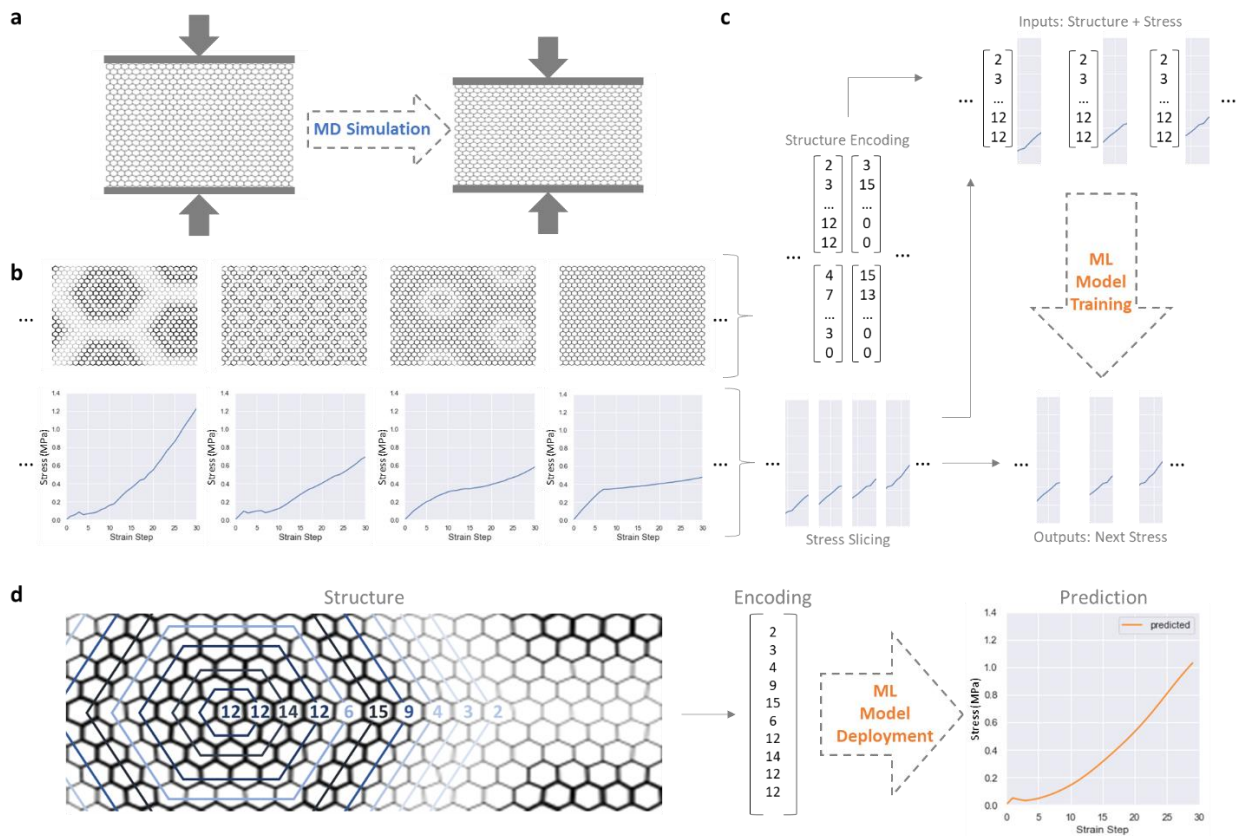

**Figure 1.** Data representation of MD simulations. **a.** MD simulation of honeycomb structures allows us to predict compressive behavior. **b.** Depending on the hierarchical architecture of the super-honeycomb, different stress behaviors result. **c.** We subsequently encode different super-honeycomb lattices into vectors representing structure, and slice stress curves into progressive windows of stress evolution. In doing so, we train an ML model to predict stress curves given an initial structure and stress condition. **d.** Once trained, the model can be deployed on a hierarchical honeycomb structure, parametrized by the cell wall thicknesses of each hierarchical hexagonal ring, at an initial unstressed state to predict the full stress-strain behavior.

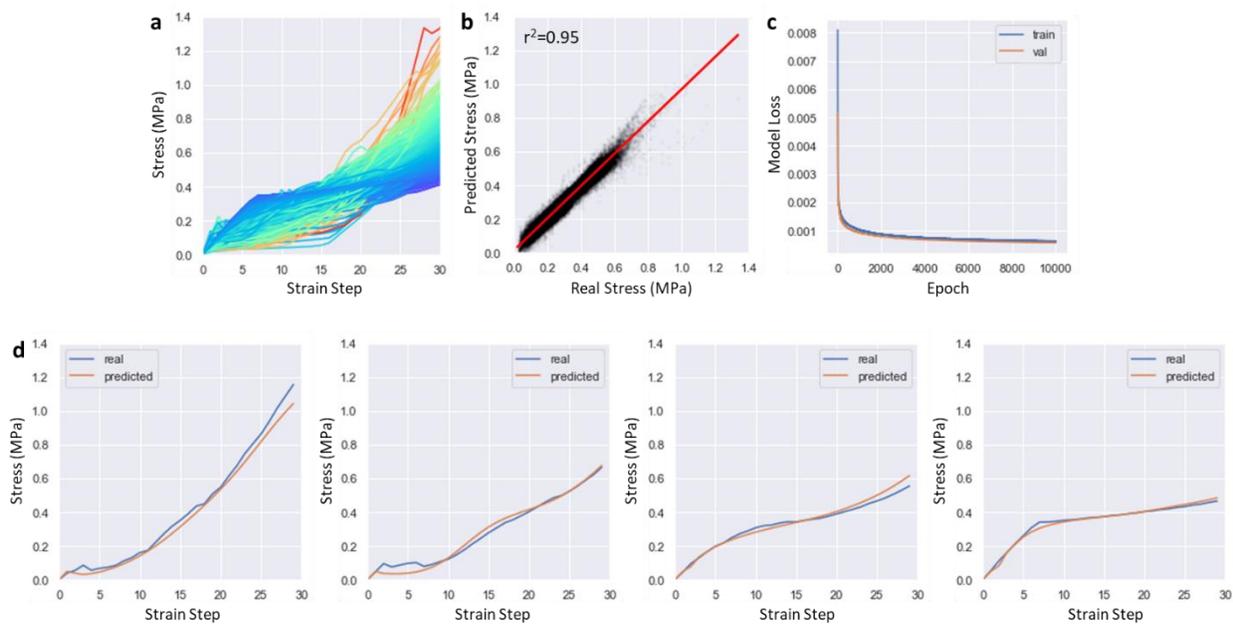

**Figure 2.** LSTM model training. **a.** An ensemble of 1,445 MD simulations were used to train the convolutional LSTM network. **b.** Predicted ML stresses align well with real MD stresses, with an $r^2$ = 0.95 and **c.** validation loss = 0.00058. **d.** Predicted curves across a range of stress behaviors align well with MD, with the samples from Figure 1b provided as example.

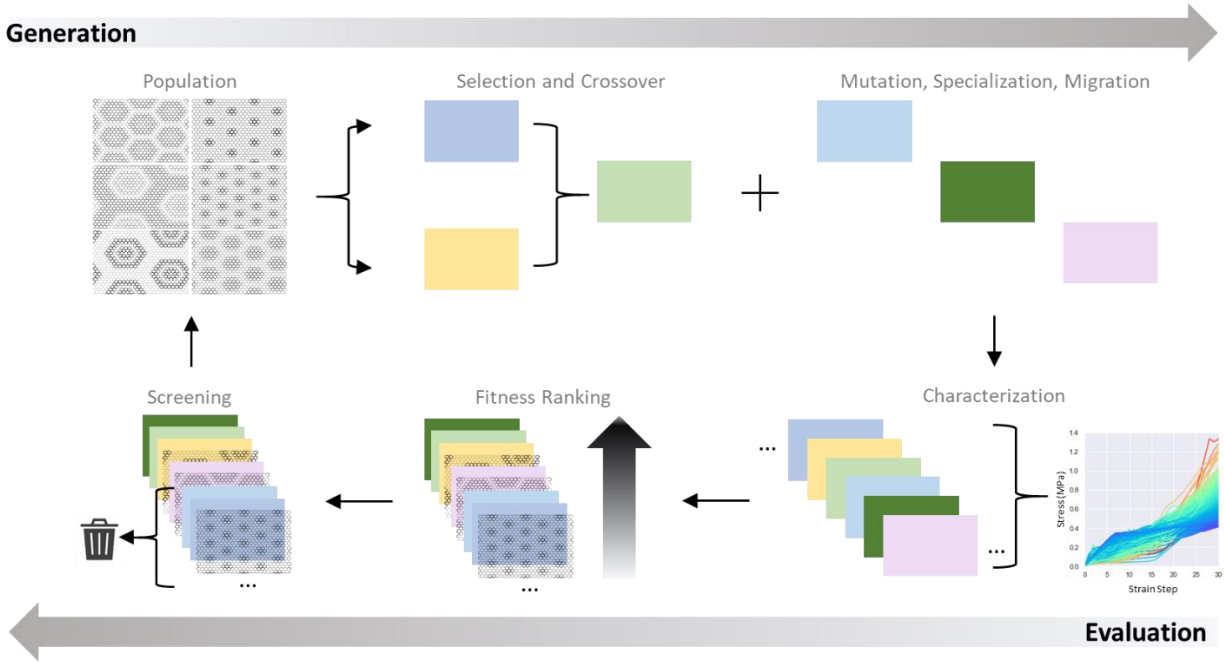

**Figure 3.** Inverse design procedure. The stress prediction ML model directly solves the forward design problem, where we input an arbitrary structure vector and rapidly receive its stress strain curve. Here, we solve the inverse design problem via genetic algorithm, which comprises an iterative two stage process of generation and evaluation, to obtain structures given a desired stress behavior as input.

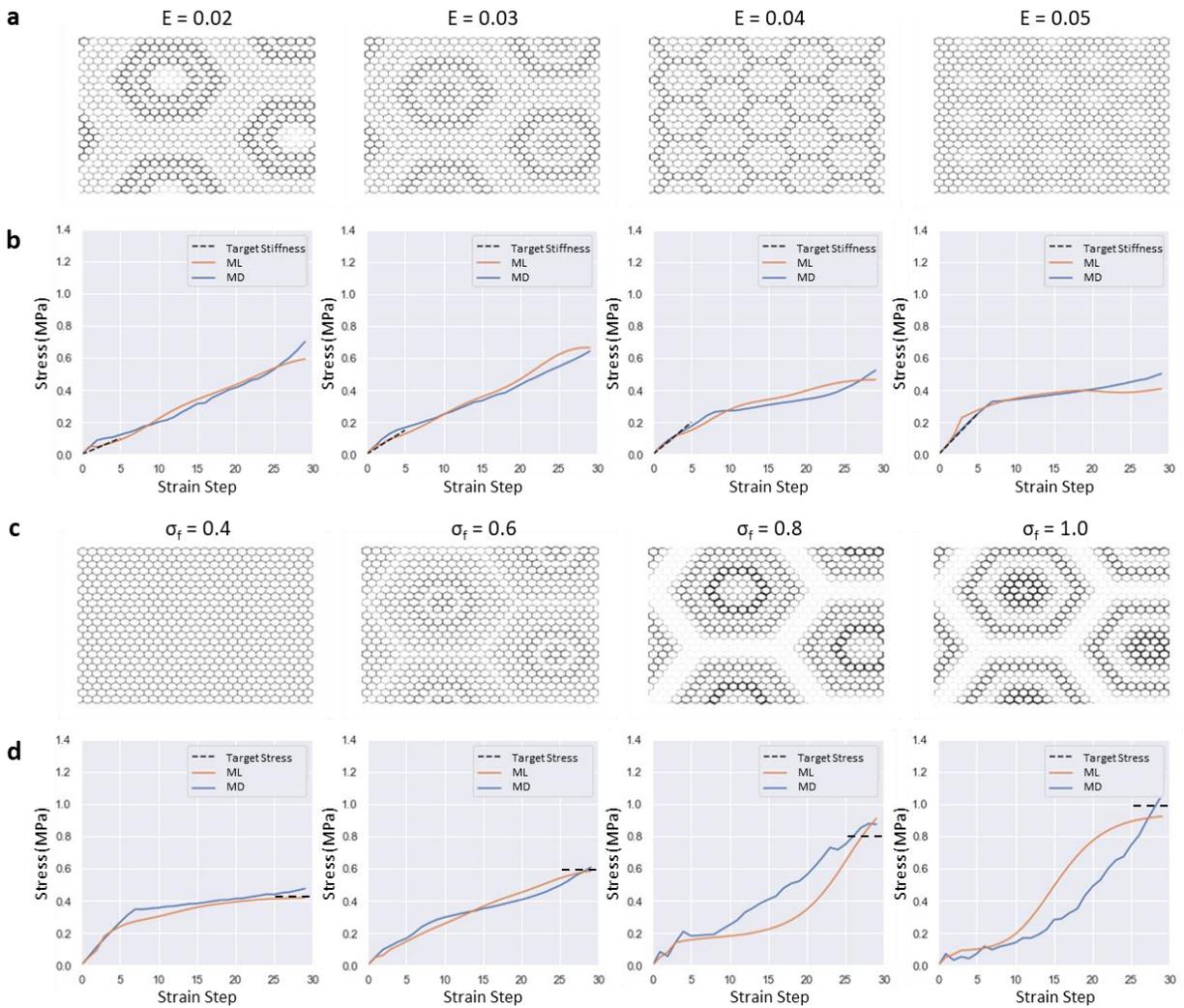

**Figure 4.** Inverse design of stiffness and ultimate stress. **a.** We input a series of desired stiffness values, and the approach generates a diversity of super-honeycomb structures predicted to meet these criteria. **b.** These AI-generated structures are subsequently confirmed to have the desired behavior by MD simulation. **c.** We do the same for a different property, such as ultimate stress, and again **d.** confirm desired progression of structures with increasing stress by MD simulation.

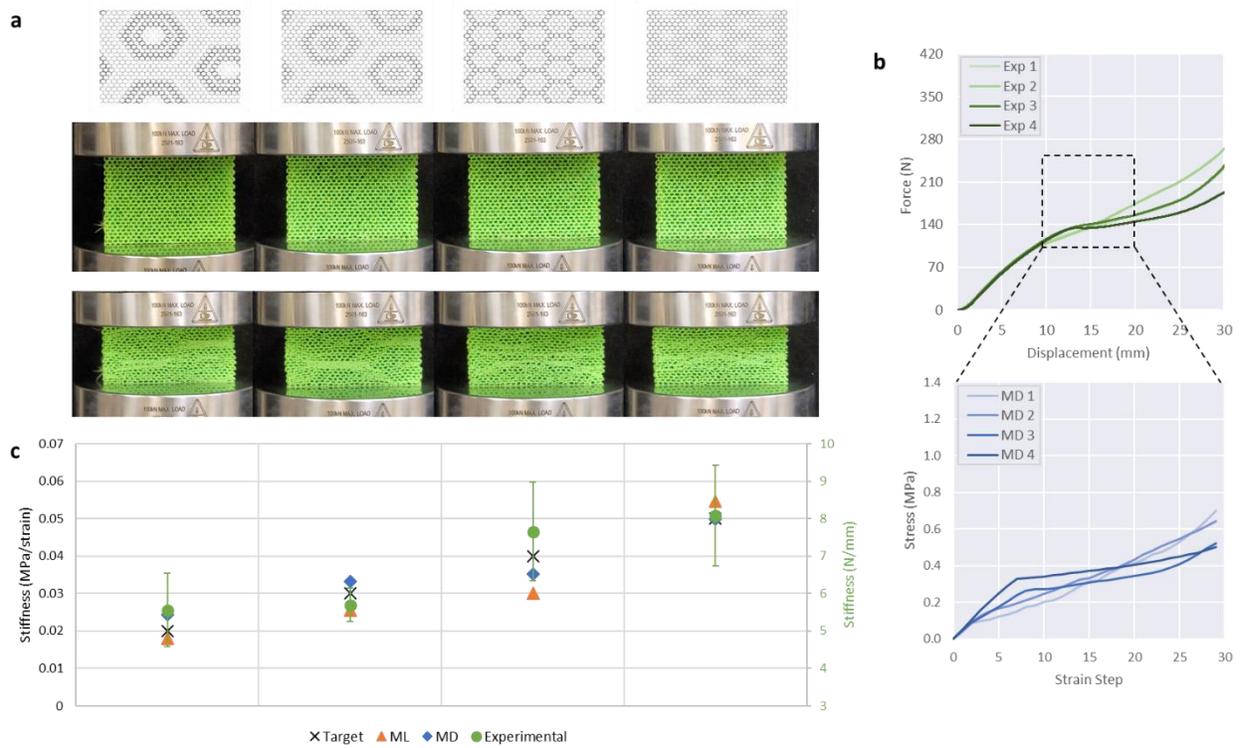

**Figure 5.** Experimental verification of stiffness design. **a.** We 3D print the AI-generated candidate structures and subject them to experimental compression tests. **b.** All samples have similar initial force displacement curves, but differences in superlattice architecture lead to different buckling responses after 10 mm. In this 10-20 mm region, compressive behavior qualitatively corresponds to simulated stress curves. **c.** Quantitatively, the relative experimental stiffnesses measured from this divergence point align well with the target behavior, ML predictions, and MD simulation.

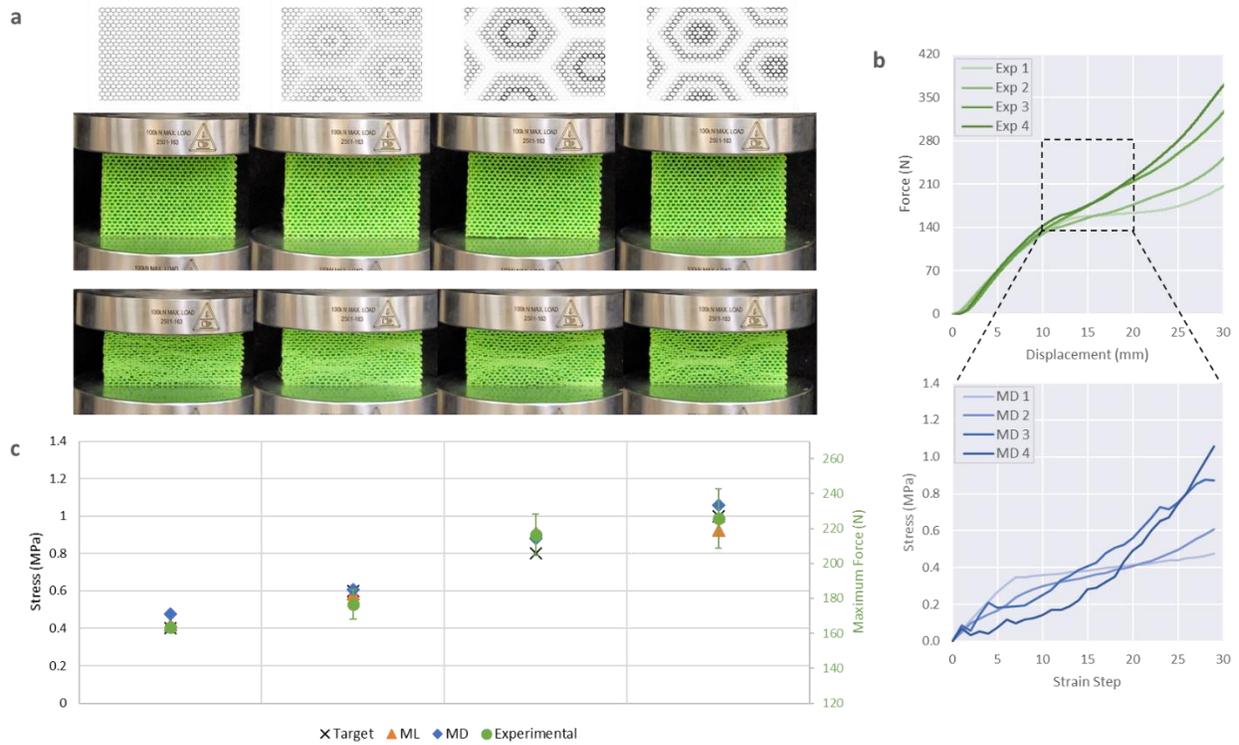

**Figure 6.** Experimental verification of stress design. **a.** We 3D print the AI-generated candidates for stress design and subject them to experimental compression tests, which **b**. again correspond to simulation in the 10-20 mm range and **c.** quantitatively align to the desired targets, ML, and MD.